\documentclass[12pt]{article}
\usepackage{authblk}
\usepackage[margin=1in]{geometry} 
\usepackage{setspace}
\usepackage{enumerate}
\usepackage{graphicx}
\usepackage{amssymb}
\usepackage{epstopdf}
\usepackage{amsmath}
\usepackage{url}
\usepackage{pdflscape}
\usepackage[utf8]{inputenc}
\usepackage{textgreek}
\usepackage{afterpage}
\usepackage[ruled,vlined]{algorithm2e}
\usepackage{lipsum}
\usepackage{array}
\usepackage{longtable}
\usepackage[labelfont=bf]{caption}
\usepackage[utf8]{inputenc}
\usepackage{hyperref}
\usepackage{blindtext}
\usepackage{comment}
\usepackage{amsmath}
\usepackage{adjustbox}
\usepackage{lscape}
\usepackage{calc}
\usepackage{float}
\usepackage{placeins}
\usepackage{makecell}
\usepackage{tabularx}
\usepackage{siunitx}
\usepackage{booktabs}
\usepackage{bm}
\usepackage{color, colortbl}
\definecolor{Gray}{gray}{0.9}
\usepackage{setspace}
\usepackage[numbers,sort&compress]{natbib}

\usepackage{amsthm}

\author[1]{\textit{Jomarie Jim\'enez Gonz\'alez}}
\author[2]{\textit{Hernando Mattei}}
\author[1]{\textit{Luis R. Pericchi Guerra}}
\author[3]{\textit{Ang\'elica M. Rosario Santos}}
\affil[1]{Department of Mathematics, University of Puerto Rico at R\'io Piedras}
\affil[2]{Department of Social Sciences, University of Puerto Rico, Medical Sciences Campus}
\affil[3]{Department of Biostatistics and Epidemiology, University of Puerto Rico, Medical Sciences Campus}

\begin{document}

\title{Methodological insights in Bayesian Age-Period-Cohort analysis: an application to the case of Puerto Rico's fertility decline}

\date{\today}
\maketitle

\section*{Abstract}
\doublespacing

Age-Period-Cohort (APC) models are of special importance in Demography and Epidemiology for analyzing panel data according to three different factors: biological (age), technological (period) and cultural (cohort). The main goal of APC modeling is to separate the explanation of both period and cohort effects to the phenomenon. The objective of this paper is to develop a Bayesian Age-Period-Cohort framework that can model a wide range of demographic and epidemiological phenomena and improve upon existing statistical methodologies. The APC framework consists of addressing three main challenges: (1) the identification problem of all APC models, usually managed by imposing constraints on effect groups, (2) considering expert knowledge in the model definition, and (3) efficient solution of computational issues. By allowing full parameter uncertainty, use of robust priors, and an efficient computational implementation, a Bayesian methodology manages these concerns. Bayesian models also produce results that allow intuitive implementation and support theoretical knowledge. Our original methodology consists of the use of  (i) a Scaled Beta2 prior distribution for the scale parameters, (ii) imposing different period and cohort constraints and comparing them,(iii) user-friendly implementation that can be easily adapted to the event, and (iv) various model comparison criteria that leads to reasonable interpretation of APC effects. We examine the dramatic collapse of fertility in Puerto Rico, an application that is difficult to model due to the accelerated changes and has interesting demographic implications that challenge the predominance of period effects in lowest-low fertility countries, emphasizing the cohort (cultural) momentum. The scope of the methodology introduced here is wide, including applications to obesity or smoking studies, for example.

\section*{Introduction}\label{Introduction}
Fertility is one of the drivers of population growth, and its understanding is of crucial relevance in Demography. Despite previous studies in the Age-Period-Cohort (APC) that highlight the importance of period effects to explain fertility decline \citep{pullum1980, billari2023age, kye2012}, our results suggest a different pattern in Puerto Rico. The APC method is powerful in analyzing panel data but requires managing the identification problem. When implementing our Bayesian probabilistic methods, which consider the identification problem in an innovative way, we find that cohort effects seem to have greater weight when describing fertility in Puerto Rico, particularly for women born in the 1963–1967 cohort and onward. These findings imply that public policies which address fertility in Puerto Rico could be most suitable when considering social and cultural values. 

By showing whether period or cohort effects are more important, APC models help reach a greater understanding of demographic processes. Models that are developed with an APC framework show how the event of interest changes for each effect and can help forecast future rates or occurrences. Age effects refer to all biological processes that occur throughout a person’s lifetime and can be described as changes at an individual level. Period effects include events that affect people of all age groups simultaneously in a specific time interval, summarized as technological effects. Economic events, wars, natural disasters, and advancements in medicine are examples of period effects \citep{demo1982review}. Cohort effects encompass all experiences shared by people born at a specific point in time, occurring as they age. These effects refer to cultural changes. Cohorts have been considered drivers of social change \citep{ryder1965} and consequently prioritized in a theoretical context. 

Incorporating APC models helps to understand the possible reasons behind the difference in total fertility rates (TFR) among countries. The TFR, calculated using age specific fertility rates (ASFR), is defined as the average number of children a woman will have in her lifetime, assuming the conditions in the analyzed time period remain constant. Fertility levels have been established to categorize countries according to their TFR. Countries above the replacement level of 2.1 have high fertility, while low fertility occurs in countries with a TFR below 2.1, which can be classified into further subcategories. Moderately low fertility corresponds to rates in the 1.7–2.0 range, while rates of 1.5 or lower define countries that have very low fertility \citep{mcdonald2002}. Countries with lowest-low fertility are defined as having a TFR of 1.3 or lower \citep{kohler2002}. 

The fertility decline in recent decades has been a matter of concern for many countries. When using data from 204 countries, the global Total Fertility Rate (TFR) is projected to be 1.83 (95\% UI 1·59–-2·08) in 2050, being lowest in South Asia with 1.36 (1·09–-1·64), and only 49 countries surpassing the replacement level of 2.1 \citep{gbd2024}. In the United States, with a  TFR of 1.6 in 2023 (United Nations, 2024), it is predicted that, compared to their predecessors, younger cohorts will have fewer children, as their total intended parity has slightly decreased \citep{hartnett2020}. Chinese fertility is expected to decrease despite measures such as the three-child policy introduced in 2021, and its decline is mostly attributed to the previous one-child policy, and the tempo effect due to postponement of marriages \citep{lan2021, yang2022china}.

In Italy, young college-educated women born in the 1960s postponed births and recovery began in the early 2000s once they had children at their thirties. In the present day, low fertility still remains in Italy due to postponement and low recovery in younger cohorts \citep{caltabiano2016}. 

A robust analysis of fertility changes should not be based solely on the TFR, since declines of this measure do not always share a common cause, such as postponement of births. Fertility postponement happens when women decide to have children at a later age, or delay the birth of their next child. Postponement can occur due to pursuing secondary education, labor force participation, economic uncertainty, and shifts in values and attitudes \citep{mills2011, vanwijk2024}. Women may wait to have children until they have completed their educational goals, have reached a stable period in their career, or until they can purchase a house, among many other reasons. Postponement can have a positive effect in the form of recovery or recuperation, in which older women have enough children to make up for lost births in earlier years \citep{fallesen2023}. Negative consequences include low fecundity and families of smaller sizes, contributing to fertility decline \citep{schmidt2011}. According to our research, Puerto Rico has not experienced a postponement of births until now. This motivates the use of fertility measures additional to the TFR, such as cumulative cohort fertility rates \citep{frejka2011role}.

 Fertility analysis of South Korea through APC models showed that period effects have a greater contribution to changes in fertility with a decline heavily influenced by family planning programs and economic development, as well as delayed childbearing \citep{kye2012}. APC analysis for fertility in the United States also attributed importance to period changes, and considered them as drivers for the baby boom and baby bust \citep{billari2023age}. In contrast with these studies, our methodologies allow obtaining a full uncertainty of the model parameters, impose constraints with specific considerations, and introduce additional model comparison criteria. The main purpose of this research is to investigate APC effects for Puerto from 1948--2022 with a Bayesian framework, to decide whether period or cohort effects explain fertility decline better. It is pertinent to consider the case of Puerto Rico, as it is currently among the countries with the lowest TFR \citep{Roser_2024}.

\subsection*{Fertility trends in Puerto Rico}

\begin{figure}[H]

\begin{flushleft} 
\centering
\caption{Total Fertility Rate in Puerto Rico for 5-year periods, 1948--2022.} 
\label{tfr-plot}
\vspace{0em}
\includegraphics[width=0.6\textwidth]{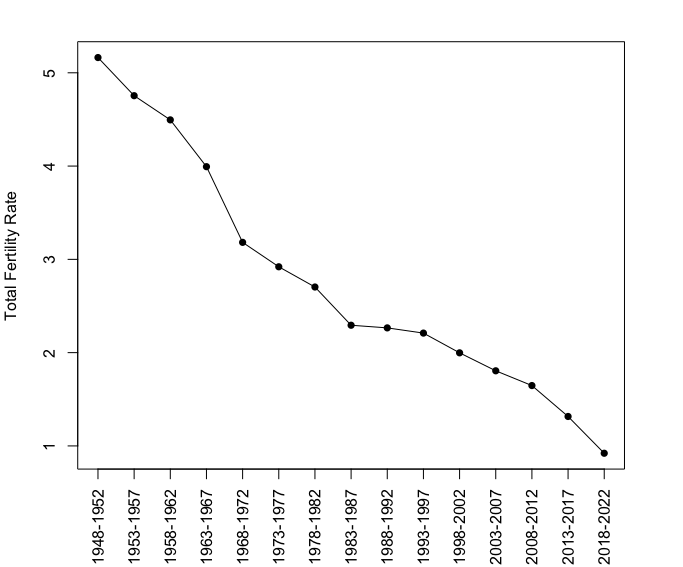}
\end{flushleft}
\end{figure}

Puerto Rico's decline in TFR is considered one of the steepest in the 21st century \citep{gbd2024}. Probabilistic projections suggest that Puerto Rico could have a TFR of 1.1  by 2050 \citep{rosario2024bayesian}. As seen in Figure \ref{tfr-plot}, 
the Total Fertility Rate (TFR) was 5.2 in the 1948-1952 period, but eventually the replacement level of 2.1 was reached in 1998-2002, representing a 61\% decrease.
This decrease has continued ever since, with a TFR of 0.9 in 2023, making it the second-lowest TFR in the American continent, and the third lowest in the world \citep{Roser_2024}. A 82\% decrease in the TFR was observed in the 2018-2022 period, when compared to 1948-1952.  

\subsubsection*{Reproductive intentions survey data}

Most recent data on reproductive intentions in Puerto Rico were collected in the 1995-96 Reproductive Health Survey (ESR1996) and the 2024 Reproductive Intentions Survey (EIR 2024) The ESR was a household survey and the 2024 Reproductive Intentions Survey (EIR 2024) was a telephone survey \citep{davila1998puerto, upr_demografia}. The EIR 2024 study consisted of a telephone survey conducted among 1,147 adults aged 18–44 across Puerto Rico, comprising 812 women and 335 men \citep{upr_demografia}. The instrument captured demographic characteristics, parity, and ideal family size, enabling comparisons with the prior ESR1994/95–96 survey, which are discussed in \cite{encuesta}. Results of both surveys show that completed family size has fallen from 2.9 in 1995-94 to 1.5 in 2024. Table 1 shows the average number of children that women in Puerto Rico have, according to the woman’s age group. Women in age groups 30-34, 35-39 and 40-44 had an average of two children in the 1995-96 period, but in 2024 that average decreased to one. 

\begin{table}[ht]
\centering
\caption{\label{tab:survey1} Average Children per Woman: ESR 1994 (1995–96) vs. EIR 2024.}
\begin{tabular}{lclc}
\toprule
\multicolumn{2}{c}{ESR 1995} & \multicolumn{2}{c}{EIR 2024} \\
\cmidrule(l{3pt}r{3pt}){1-2} \cmidrule(l{3pt}r{3pt}){3-4}
Age Group & Average & Age Group & Average\\
\midrule
15-19 & 0.16 &       & \\
20-24 & 0.81 & 18-24 & 0.34\\
25-29 & 1.62 & 25-29 & 0.72\\
30-34 & 2.16 & 30-34 & 1.02\\
35-39 & 2.44 & 35-39 & 1.34\\
40-44 & 2.53 & 40-44 & 1.48\\
\addlinespace
45-49 & 2.89 &        & \\
\bottomrule
\end{tabular}
\end{table}

The percentage of women in Puerto Rico without children according to both surveys is shown in Table \ref{tab:survey2}. The most notable changes are among women aged 25-29, where the percentage of childless women increased by 35.46\%, followed by the 30-34 age group with a 27.11\% increase. According to the 2024 survey, almost a third of Puerto Rican women aged 40-44 have no children. 

\begin{table}[ht]
\centering
\caption{\label{tab:survey2} Percentage of Women Without Children: ESR 1994 (1995–96) vs. EIR 2024.}
\begin{tabular}{lclc}
\toprule
\multicolumn{2}{c}{ESR 1995} & \multicolumn{2}{c}{EIR 2024} \\
\cmidrule(l{3pt}r{3pt}){1-2} \cmidrule(l{3pt}r{3pt}){3-4}
Age Group & Average & Age Group & Average\\
\midrule
15-19 & 86.57 &  & \\
20-24 & 53.50 & 18-24 & 78.40\\
25-29 & 26.14 & 25-29 & 61.60\\
30-34 & 14.49 & 30-34 & 41.60\\
35-39 & 10.49 & 35-39 & 35.60\\
40-44 & 9.67 & 40-44 & 27.20\\
\addlinespace
45-49 & 8.24 &  & \\
\bottomrule
\end{tabular}
\end{table}

From the EIR 2024 survey the lack of a classic birth postponement pattern is evident: 85\% of mothers near the end of the reproductive span (aged 40-44) report no plans of having more children, and approximately one-quarter of childless women do not desire motherhood. These strong intentions against childbearing at later ages are consistent with a cohort-driven, low-quantum regime rather than delayed timing alone. 

\subsection*{Demographic indicators of Puerto Rico}

The decline in TFR, as well as negative net migration due to low immigration of foreigners to Puerto Rico, led to a population increase observed until 2004, that is attributed to demographic momentum. The population of Puerto Rican women in reproductive age began to decrease in 1998, and though the TFR reached 2.1 in 1998, it is not until 2005 that the general population began to decrease.
The combination of Puerto Rico's current demographic indicators: negative natural growth, a TFR below 1.0, negative net migration and a high proportion of older adults create a unique situation that is not observed in other populations of the world. As in other countries, the decline in Puerto Rican fertility is associated with population aging. The age composition of Puerto Rico has changed notably since 2004, where the 0--14 age group represented 24\% of the population, the 15--64 age group 66\% and the 65+ age group 12\%. This contrasts with the population structure in 2023, consisting of 13\% in the 0--14 age group, 64\% in the 15--64 age group, and 24\% in the 65+ age group, which now classifies the population in Puerto Rico as super-aged.

Puerto Rico experiences a negative net migration, which is due to the low number of immigrants that arrive each year. The frequency of interstate migrants is very low in Puerto Rico, when compared to the 50 US states and Washington DC. Puerto Rico is the territory with the lowest fertility rate and the lowest immigration, according to data from the Puerto Rico Community Survey (PRCS) \citep{uscensus2024prcs} and the American Community Survey (ACS) \citep{uscensus2024acs}. 

We follow Bayesian methodologies to develop a model that describes fertility in Puerto Rico from 1948 to 2022, in order to analyze age, period, and cohort effects. Puerto Rico's unique demographic situation described earlier justifies its analysis through an APC framework. The main scientific and demographic contribution is the fact that our analysis suggests the importance of cohort effects for describing fertility decline, which differs from other countries. These findings are necessary for developing appropriate public policies that address fertility changes. The main methodological contributions are the constraints imposed in the model, the use of Scaled Beta 2 priors, and additional model comparison criteria. 

\section*{Background}

The research works of Mason, Fienberg, and collaborators \citep{fienberg1979identification, mason1985, mason1973}, are the foundation of modern APC analysis. Most APC models found in literature focus on mortality \citep{caselli1989age} and incidence of chronic diseases, such as cancer \citep{clayton1987a, negri1990}, heart disease \citep{su2022}, and obesity \citep{keyes2010}. APC models have also been applied to explain migration \citep{bozick2021, sander2016}, verbal test scores \citep{yang2006mixed}, political participation \citep{grasso2019}, and religious beliefs \citep{vera2021age}.

Several models have been used to describe fertility. A notable example is the marital fertility model proposed by \cite{coale1974}, which is multiplicative for the rates and assumes natural fertility, meaning fertility is not purposely controlled \citep{henry1961}. The Bongaarts Fertility model incorporates data on contraceptive use, abortions, and infecundability \citep{bongaarts1983}. Fertility analysis using an APC approach is not as common in literature when compared to mortality analysis. Early development of APC models applied to fertility concerned the United States, \citep{pullum1980}, and the Netherlands \citep{Willekens1984}. Countries analyzed in recent APC models include the United States \citep{billari2023age}, China \citep{lan2021}, Taiwan \citep{tzeng2019trends}, South Korea \citep{kye2012}, Italy \citep{caltabiano2016}, and Japan \citep{okui2020marriage}. 

\subsection*{Bayesian Age-Period-Cohort models}
\label{bck-bayes}
 In Bayesian analysis, prior information or expert knowledge is incorporated into the model, which is then updated by the observed data. The posterior distribution of the estimated parameter $\theta$ can be obtained using the following relationship:

\vspace{-6pt}
\begin{equation}
    \pi(\theta|x) \propto f(x|\theta)\pi(\theta),
\end{equation}
where x is the data sample, $f(x|\theta)$ is the likelihood, and $\pi(\theta)$ is the prior distribution. A Bayesian framework allows a direct approximation of probabilities \citep{besag1995bayesian}, as well as a full evaluation of the uncertainty in random effects and functions of parameters \citep{breslow1993}. The probability intervals of the parameters allow for a more intuitive interpretation. The first Bayesian models are the Age-Cohort model proposed by \cite{breslow1993}, applied to breast cancer rates, a model describing deaths by prostate cancer \citep{besag1995bayesian}, and the contributions of \cite{berzuini1993bayesian} and \cite{berzuini1994bayesian}. Statistical inference is executed through Monte Carlo Markov Chain (MCMC). Further development has led to multivariate APC models that compare effects and rates across strata, such as regions in England and Wales \citep{riebler2010}, and women and men \citep{torres2017}. 

In demography, Bayesian methodologies have gained popularity in recent years. Since 2015, the United Nations World Population Prospects incorporates probabilistic projections of populations \citep{raftery2014pop}, life expectancies \citep{raftery2014joint}, and total fertility rates \citep{liu2020}. Bayesian APC models in literature emphasize forecasting cancer mortality \citep{bray2002application} and population \citep{havulinna2014}. Fertility analysis can also consider parities. Fertility rates have been projected using parametric mixture models \citep{hilton2020} and generalized additive models \citep{ellison2024}, a similar approach to APC models. 

\subsection*{The identification problem}
\label{bck-id}
APC analysis commonly relies on cross-sectional data, where the rows of the table are the different age groups and the columns represent the periods. Cohorts are obtained by the collinear relationship $cohort = period - age$. The identification problem is caused by this linear dependency, where the singular design matrix produces an infinite number of solutions, meaning we cannot distinguish between age, period, and cohort effects \citep{yanglandbook}. All APC models must be defined in a way that addresses the identification problem, whether the methodology is frequentist or Bayesian. The most procedure is imposing constraints on the model. \cite{fienberg1979identification} suggest a sum-to-zero constraint, where the sum of the age parameters equals to zero, as well as each sum of the period and cohort parameters. Another constraint often used, known as an equality constraint, consists of fixing specific subsequent parameters to zero, where, for example, the parameters for the first and second age group are equal to zero. The effects set to zero are the reference variables. This technique has limitations, as the constraints should be based on theoretical assumptions, and change the values of the coefficients. 

A popular method that deals with the identification problem is the Intrinsic Estimator (IE) \citep{yang2008intrinsic}. This is a general purpose method that modifies the vector of solutions for the parameters. The IE has characteristics that are statistically favored such as unbiasedness and consistency. Despite these advantages, some researchers argue that it lacks robustness because its results depend on the design matrix: the number of APC categories, the selected reference category and the size and sign of the nonlinearities \citep{fosse2019}. 

\subsection*{APC analysis software}
\label{bck-pkg}
Several R packages have been developed that perform APC analysis. The package \textit{apc} \citep{apc_pack} follows a frequentist framework and imposes constraints on the second-order differences of the parameters. A Bayesian alternative is proposed on the package \textit{bamp} focusing on applications to incidence and mortality \citep{bamp-pack}. It adds sum-to-zero constraints and additional linear transformations to improve identifiability. This package could not be used with our data because convergence was not reached, and the simulation had a long running time. Another alternative for implementing Bayesian APC models is to use an MCMC statistical package, such as JAGS (Gibbs Sampling), and Stan (Hamiltonian Monte Carlo Sampler). This alternative requires defining the hierarchical model in code, which provides the user with full control and customization when defining the priors. Stan, the software selected for our APC model implementation, has the advantages of reasonable simulation running time, extensive online documentation, frequent updates, and interfaces in many programming languages, such as R and Python. The specifications for implementing our model are explained in the Methods section.

\section*{Methods}\label{met}

The APC analysis uses data from the Puerto Rico Demographic Registry and the US Census Bureau. \textcolor{black}{The data is given by matrix $Y_{A \times T}$, denoting the number of births. The rows correspond to $A = 7$ age groups of women in reproductive age, where each group is of width 5: 15-19, 20-24, 25--29, 30--34, 35--39, 40--44 and 45--49. The columns indicate the $T = 15$ periods in 5-year intervals from 1948 to 2022: 1948--1952,..., 2018--2022. Matrix $P_{A \times T}$ describes the number of women of each age group and period in person-years, with the same structure used to define $Y$}. From matrix $Y$ we can calculate the ASFR for each entry, and subsequently obtain the TFR of all periods. The cohorts are found in the diagonals of the table, with the total number of cohorts being $C = (A-1) + T = 21$. 

\subsection*{Bayesian model definition}

Let $y_{a,t}$ be the number of births for age group $a$ in a time period $t$, defined with a Poisson likelihood:

\vspace{-6pt}
\begin{equation}
 y_{a,t} \sim \text{Pois}(\lambda_{a,t})
\end{equation}
\vspace{-6pt}

\vspace{-6pt}
\begin{equation}\label{eqn:lambda_eq}
 \log(\lambda_{a,t}) = \lambda_0 + \theta_a + \phi_t + \alpha_c - \log(p_{a,t}),
\end{equation}
\vspace{-6pt}

where $a = 1,\dots, A, t = 1,\dots, T$ and $c = (A-a) + t$.
The parameters $\theta_a$ are the age effects, $\phi_t$ the period effects, and $\alpha_c$ the cohort effects. Note that $\lambda_0 \sim \text{N}(0, 1)$ is a measurement error, and$e^{\lambda_{a,t}}$ represents the birth rate. We subtract the offset in (\ref{eqn:lambda_eq}) to model the rates instead of the expected occurrences. This model definition is also used for Age-Period (AP), Age-Cohort (AC) and Age (A) models.

The age, period, and cohort parameters follow autoregressive priors, with linear time trends specified by a second-order Random Walk (2). This is a type of Random Walk in which the second-order differences remain constant. For the age parameters, these constant second-order increments are defined as

\vspace{-6pt}
\begin{equation}
\label{eqn:rw2}
\theta_a - \theta_{a-1} = \theta_{a-1} - \theta_{a-2}
\end{equation}
\vspace{-6pt}

In particular, the priors for the age parameters $\theta_a, a = 1, \dots, A$ were formulated as follows:

\vspace{-6pt}
\begin{equation}
\label{eqn:age-eff}
\theta_a \sim \text{N}(2\theta_{a-1} - \theta_{a-2}, \tau_1^{-1})
\end{equation}
\vspace{-6pt}

Note that the mean of the normal prior in Equation \ref{eqn:age-eff} is obtained by solving for $\theta_a$  in Equation \ref{eqn:rw2}.

The same structure is used for defining the priors in the period parameters:

\vspace{-6pt}

\begin{equation}
\phi_t \sim \text{N}(2\phi_{t-1} - \phi_{t-2}, \tau_2^{-1})
\end{equation}
\vspace{-6pt}

And for the cohort parameters:

\vspace{-6pt}
\begin{equation}
\alpha_c \sim \text{N}(2\alpha_{c-1} - \alpha_{c-2}, \tau_3^{-1})
\end{equation}
\vspace{-6pt}

The precision parameters $\tau_j, \ j = 1,2,3$ are assigned a Scaled Beta2 distribution \citep{perez2017scaled}, which depends on the hyperparameters $p$, $q$, and $b$:

\vspace{-6pt}
\begin{equation}
  \text{SBeta2}(\tau_j | p, q, b) = \frac{\Gamma(p+q)}{\Gamma(p)\Gamma(q)\cdot b} \cdot \frac{(\frac{\tau_j}{b})^{(p-1)}}{((\frac{\tau_j}{b})+1)^{(p+q)}} 
\end{equation}
\vspace{-6pt}

The advantages of the Scaled Beta 2 distribution include its flexibility, meaning that it allows for heavier tails depending on the prior definition, and its robustness (less sensitive to extreme values).  The half-Cauchy distribution, which has been proposed as an alternative to the inconvenient Gamma prior often used in literature \citep{gelman2006prior}, is a special case of the Scaled Beta2 distribution. It is also equivalent to an F distribution, for the case of $b = \left(\frac{q}{p}\right)^p$. The elicitation of the Scaled Beta2 distribution hyperparameters, as well as the relation between the F distribution and the Scaled Beta2, warrants a detailed discussion, and will therefore be further explored in a future paper\citep{jimenez2026}. The parameters for the precision priors were defined as $\tau_j \sim \text{SBeta2}(0.5, 0.5, 100)$. The following constraints were imposed:

\vspace{-6pt}
\begin{align}
    \theta_1 &\sim \text{N}(0, 0.01) \label{eqn:age_const}\\ 
    \phi_1 & = 0 \\
    \alpha_{10} & \sim \text{N}(0, 0.0025) \label{eqn:coh_const1}\\ 
    \alpha_{11} &\sim \text{N}(0, 0.0025)  \label{eqn:coh_const2}
\end{align}
\vspace{-6pt}

Since the period constraint was imposed for the first period effect (1948-1952), we set constraints for the 10th cohort that refers to the same time interval, and for the 11th cohort (1953-1957). This cohort constraint is more suitable, because complete fertility data is available for the 10th and 11th cohort groups. The parameters $\theta_1$, $\alpha_{10}$ and $\alpha_{11}$ were initially set to exactly zero. This produced unreasonably narrow credible intervals for the age and cohort effect parameters. These narrow intervals persisted even with different prior hyperparameters. To address this issue, we added uncertainty when imposing constraints, as shown in equations 8, 10, and 11. This alternative to imposing constraints allows obtaining credible intervals of reasonable width. 

APC analysis is performed using the statistical modeling platform Stan \citep{carpenter2017stan}, through the R package \textit{rstan} \citep{rstan}, that achieves Bayesian inference with Markov Chain Monte Carlo (MCMC) methods. All models were run for 15,000 iterations. To improve convergence, centering was done on the age, period and cohort parameters.

\subsection*{Model comparison criteria}

To compare all models, we will use the Approximate Leave-One-Out Information Criterion (LOOIC) and the Widely Applicable Information Criterion (WAIC). Both criteria measure the out-of-sample prediction accuracy of the models \citep{vehtari2017}. The most suitable model has the lowest LOOIC or WAIC value.  We will calculate the LOOIC and WAIC for models with different Scaled Beta2 parameters, and a Gamma prior for the precision, in order to observe how the choice of prior affects the model's performance. The LOOIC and WAIC will also be used to compare models according to the type of constraints imposed. The constraints described in the previous subsection will be compared to the sum to zero constraint often found in literature, where $\sum_a\theta_a = \sum_t\phi_t =\sum_c \alpha_c =0$. Models will be compared using the residual errors for each model.  The residual sum of squares will be calculated for each model, using the observed ASFRs as data $\frac{y_{a,t}}{p_{a,t}}$ and the ASFRs given by each model as the fitted values.

\begin{align}
APC^2 &= \sum_i e^2_{APC, i}  & AP^2 &= \sum_i e^2_{AP, i} \nonumber \\
AC^2  &= \sum_i e^2_{AC, i}   & A^2  &= \sum_i e^2_{A, i}
\end{align}

Analogous to the coefficient of determination ($R^2$), the proportion of variation explained by the variables is calculated as such:

\begin{align}
C_{APC} &= \frac{AP^2-APC^2}{AP^2} &  P_{APC} &= \frac{AC^2-APC^2}{AC^2} \nonumber \\
PC_{APC} &= \frac{A^2-APC^2}{A^2} &  P_{AP} &= \frac{A^2-AP^2}{A^2} \nonumber \\
C_{AC} &= \frac{A^2-AC^2}{A^2},
\end{align}

where $C_{APC}$ is the proportion of the APC model variation explained by cohort effects, $P_{APC}$ is the proportion of the APC model explained by period effects, $PC_{APC}$ denotes the proportion of the APC model attributed to both period and cohort effects, $P_{AP}$ shows the proportion of variation explained by period effects in the AP model, and $C_{AC}$ is the proportion of variation in the AC model explained by cohort effects. To evaluate the ability of the APC model in producing estimates, we will also run our model using data of the first 14 periods. The model then estimates the TFR for the last period, corresponding to 2018-2022. To observe how the choice of prior affects the estimated values, this model will be developed for different prior types.

\section*{Results}

\subsection*{Exploratory Analysis}

\begin{figure}[H]
\begin{flushleft} 
\caption{Age Specific Fertility Rates in Puerto Rico, from 1950-2022.} 
\label{asfr}
\vspace{0em}
\includegraphics[width=1.0\textwidth]{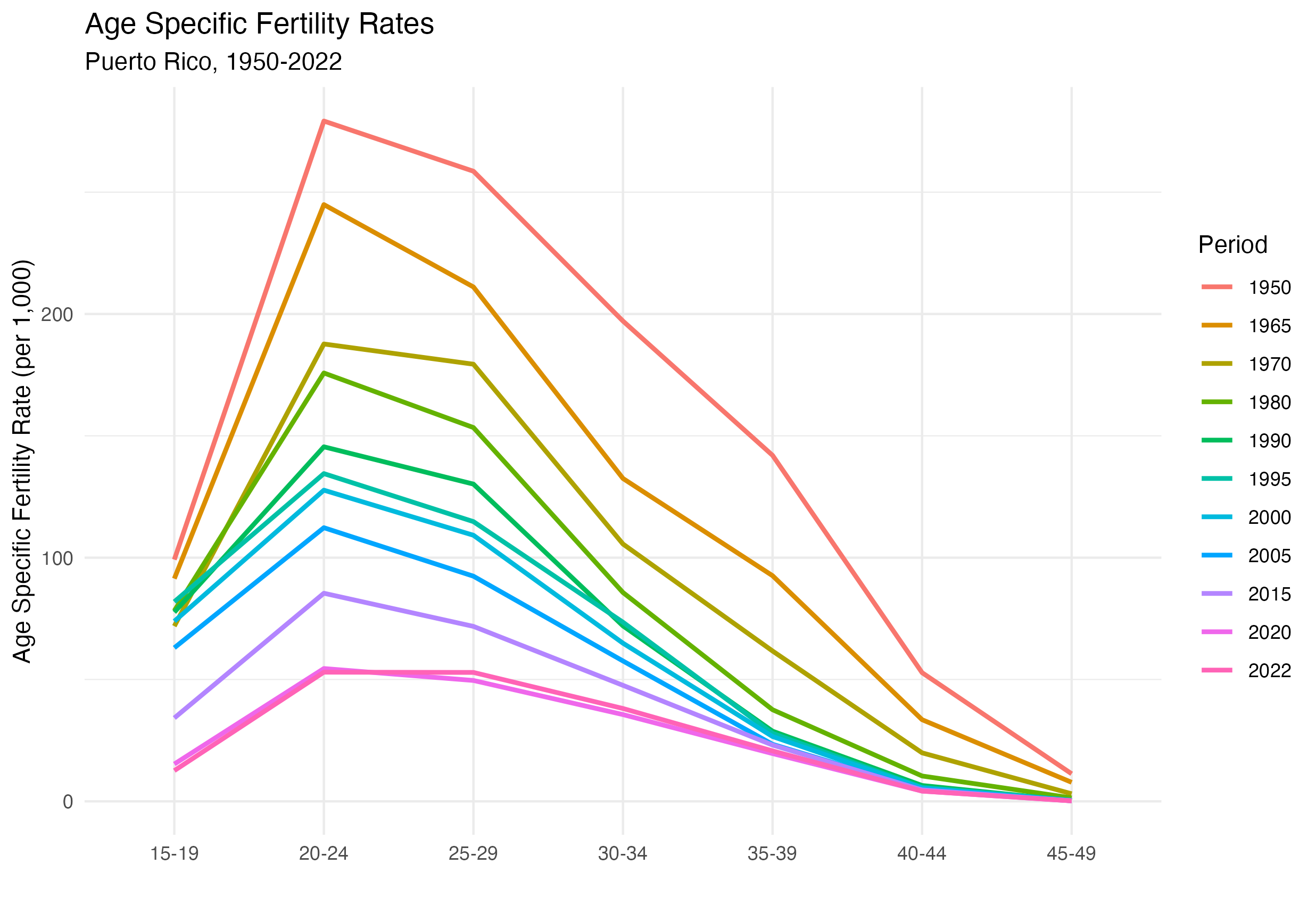}
\end{flushleft}
\end{figure}

Figure \ref{asfr} shows the Age Specific Fertility Rates (ASFR) shown for specific single-year periods. The following fertility pattern remains constant in all periods: The 20-24 age group contributes the most to fertility. The rates decrease for older age groups, suggesting the lack of postponement of births.

\begin{figure}[H]
\begin{flushleft} 
\centering
\caption{Hexamap of Age Specific Fertility Rates in Puerto Rico.} 
\label{hexamap}
\vspace{0em}
\includegraphics[width=1.1\textwidth]{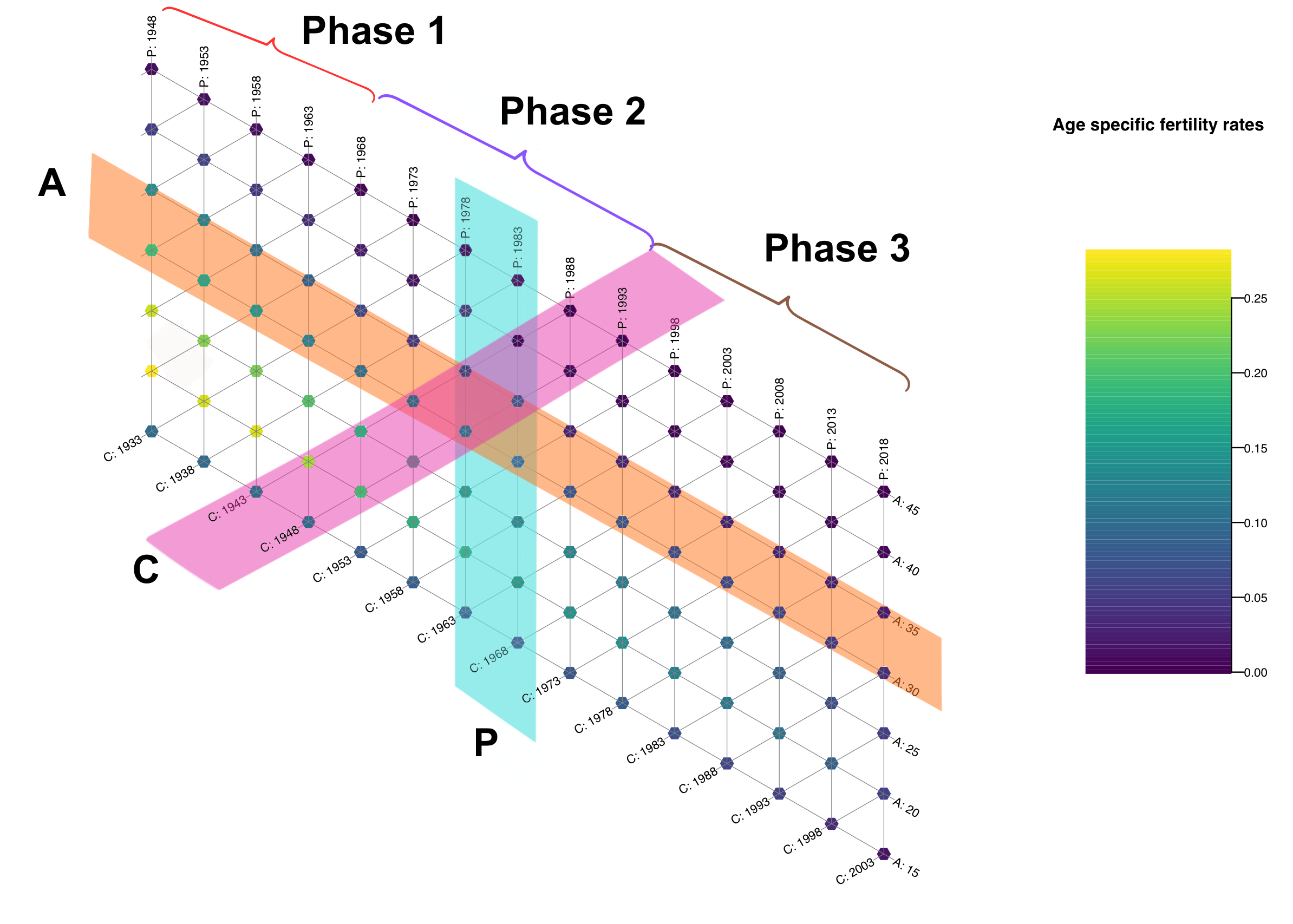}
\end{flushleft}
\end{figure}

Figure \ref{hexamap} is a hexamap (hexagonally binned heatmap, generated through the R package \textit{APCtools}) of the ASFRs, that allows visualizing all APC dimensions at once. Cohorts are labeled beginning with the 1933-1937, since it is the oldest cohort of completed fertility in the data. Age, period, and cohort groups are highlighted in the figure for clarity. According to this hexamap, fertility in Puerto Rico can be divided into three phases. In the first phase, which includes the periods 1948-1952, 1953-1957 and 1958-1962, the women who mainly contribute to fertility belong to the 20-24, 25-29 and 30-34 age groups.

Fertility rates remain stable in the 20-24 and 25-29 groups, while they decline for the 30-34 group. The 30-34 age group has an approximate contribution of 15\% to total fertility in these three periods. The second phase, which begins in the period 1963-1967 and ends in the period 1988-1992, shows that the rates decrease even for young women. This phase is dominated by the period effect. Second-order fertility rates decline in the 1980s, so that they are no longer similar to first-order fertility rates. First-order births also decline rapidly in this decade. Women aged 30-34 do not contribute much to fertility, and in the mid-1980s, it is observed for the first time that in this age group most births are second-order, rather than third-order. In the third phase, it is observed that in the period 1993-1997 onward the rates remain low, following the same pattern in the age groups. The differences in rates are not very noticeable between the periods, indicating that the cohort effect predominates. The cohorts in this phase follow a pattern of low fertility, with rates decreasing in women aged 20-24 and 25-29 years. 

\newpage
\subsection*{Bayesian APC Model effect estimates}

\begin{figure}[H]
\begin{flushleft} 
\centering
\caption{Effects of the Bayesian APC model, with 95\% credible intervals} 
\label{eff-plt}
\vspace{0em}
\includegraphics[width=1.0\textwidth]{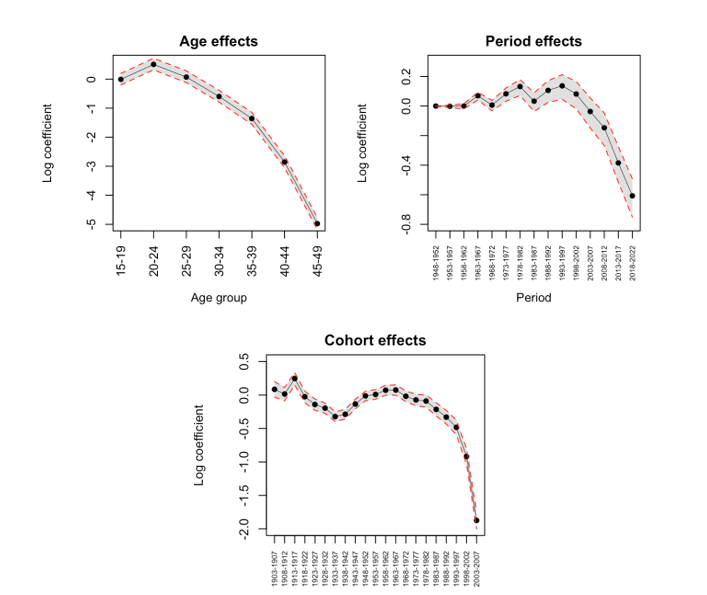}
\end{flushleft}
\end{figure}

 All presented models reached convergence, with no divergent transitions and $\hat{R} \approx 1$ for all parameters. The estimated effects of our Bayesian APC model are plotted in Figure \ref{eff-plt}. The age effects have an inverted U-shape, with births being most common in age group 20-24. Period effects increase from 1948-1998, then drop drastically. Cohort effects begin to decrease, then increase slightly for women born in 1938-1942 to women born in 1963-1967. A drastic decrease in births is seen for women born in 1978-1982 and onward.

\begin{figure}[H]
\begin{flushleft} 
\centering
\caption{Comparison of period effects in APC and AP models.} 
\label{apc-applot}
\vspace{0em}
\includegraphics[width=0.50\textwidth]{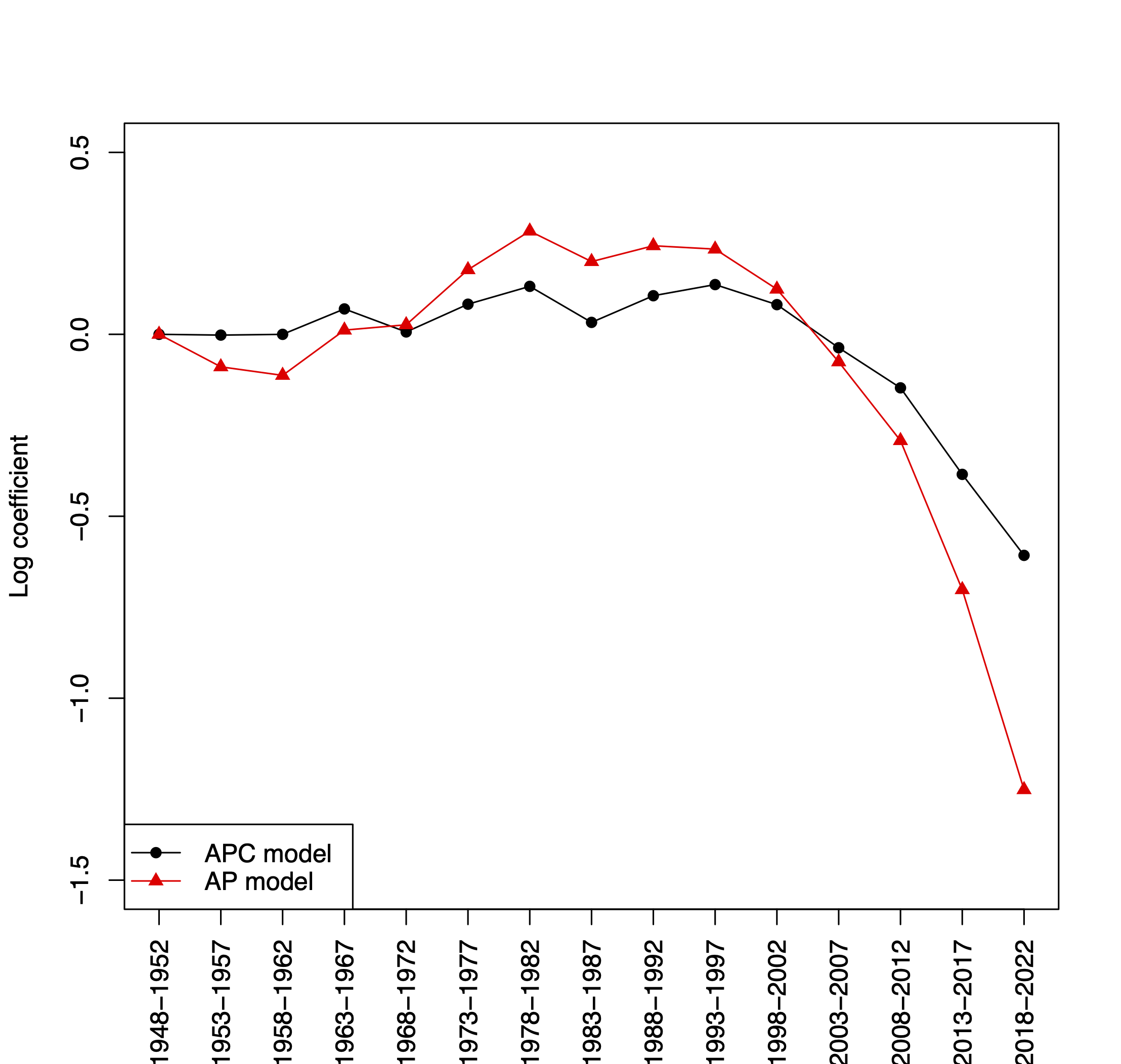}
\end{flushleft}
\end{figure}

\begin{figure}[H]
\begin{flushleft} 
\centering
\caption{Comparison of cohort effects in APC and AC models.} 
\label{apc-acplot}
\vspace{0em}
\includegraphics[width=0.5\textwidth]{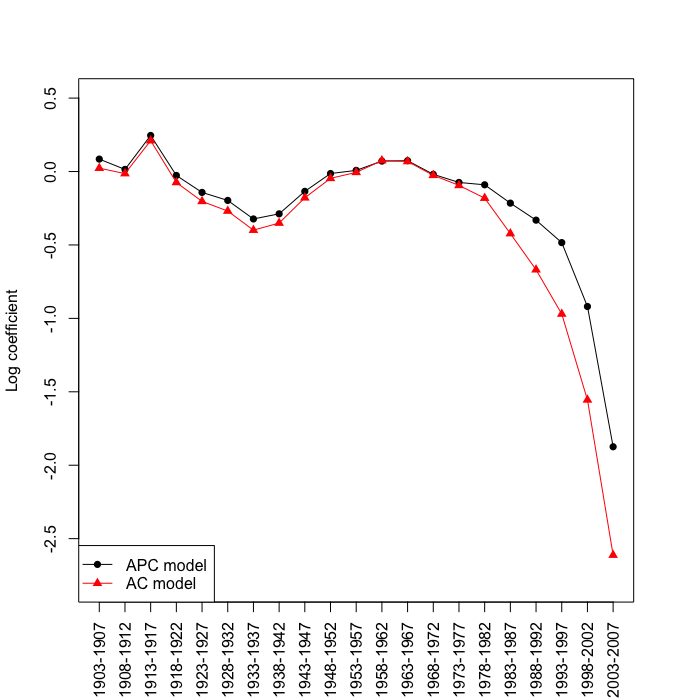}
\end{flushleft}
\end{figure}

When observing Figure \ref{apc-applot}, which compares period effects between the APC and AP models, these period effects differ in the left and right extremes. In contrast, the cohort effects shown in Figure \ref{apc-acplot}, are not that different for the APC and AC models.

\subsection*{Bayesian model comparison criteria}

Results of the residual sums of squares criterion are shown in Table \ref{tab:res}. Cohort effects explain 45.96\% of the APC model, while period effects explain 19.34\% of the model. Table \ref{tab:ic} shows the results for the LOOIC and WAIC values. The Scaled Beta2 precision priors in rows 2-5 of Table 2 are equivalent to an F distribution. The LOOIC and WAIC values do not vary drastically among priors. Both the LOOIC and the WAIC favor the APC model with a Scaled Beta 2 prior, since these models generate the lowest LOOIC and WAIC values. The LOOIC favors a SB2(2, 5, 6.25) prior, while the WAIC favors a SB2(0.5, 0.5, 100) prior.

\begin{table}[H]
\centering
\caption{\label{tab:res}Amount explained in model by period and cohort effects}\vspace{-4pt}
\begin{scriptsize}
\setlength{\tabcolsep}{4pt}
\begin{tabular}[t]{ll}
\hline
\\[-3pt]
\textbf{Formula} & \textbf{Value}\\
\\[2pt]
\hline
\\[-3pt]
$C_{APC}$  & 0.4597\\[2pt]
$P_{APC}$  & 0.1869\\[2pt]
$PC_{APC}$ & 0.9067\\[2pt]
$P_{AP}$   & 0.8273\\[2pt]
$C_{AC}$   & 0.8852
\\[3pt]
\hline
\end{tabular}
\end{scriptsize}
\end{table}

\begin{table}[H]
\centering
\vspace{-4pt}
\caption{\label{tab:ic}LOOIC and WAIC values for different models}
\begin{scriptsize}
\setlength{\tabcolsep}{4pt}

\begin{tabular}{lllll}
\hline
\\[-3pt]
\multicolumn{1}{l}{\textbf{Model}} & \multicolumn{1}{l}{\textbf{LOOIC}} & \multicolumn{1}{l}{\textbf{Standard Error}} & \multicolumn{1}{l}{\textbf{WAIC}} & \multicolumn{1}{l}{\textbf{Standard Error}}\\
 \\[2pt]
\hline
\\[-3pt]
APC, SB2(0.5, 0.5, 100) prior & 41,900.4              & 5,572.9 & \textbf{62,587.9} & 10,326.5 \\[2pt]
APC, SB2(1, 5, 5)    prior    & 41,980.1              & 5,609.1 & 62,682.0 & 10,398.6\\[2pt]
APC, SB2(2, 5, 6.25) prior    & \textbf{41,897.5}     & 5,600.1 & 62,758.4 & 10,372.8\\[2pt]
APC, SB2(1, 25, 25)  prior    & 41,919.5 & 5,591.2    & 62,681.4 & 10,341.3\\[2pt]
APC, SB2(2, 25, 156.25) prior & 41,966.0 & 5,594.6    & 62,601.0 & 10,421.9\\[2pt]
APC, G(0.001, 0.001) prior    & 41,919.8 & 5,591.9    & 62,837.0 & 10,438.8\\[2pt]
AP                            & 112,977.0 & 15,647.4  & 141,972.4 & 20,397.8\\[2pt]
AC                            &  70,734.2 & 7,925.5   & 93,565.8 & 11,604.3\\[2pt]
A                             & 566,350.3 & 141,327.4 & 618,784.7 & 155,616.6
 \\[3pt]
	\hline
\end{tabular}
\end{scriptsize}
\end{table}

We present the LOOIC and WAIC values in Table \ref{tab:const} that also compare the APC models according to parameter constraints. Both the LOOIC and WAIC values for the SB2 models suggest that our original constraints perform better. For the Gamma models, the LOOIC favors our original constraints, while the WAIC prefers a sum-to-zero constraint, and the Gamma prior overall.

\begin{table}[H]
\centering
\vspace{-4pt}
\caption{\label{tab:const}Comparison of LOOIC and WAIC values for APC models using different precision priors and effect constraints.}
\begin{scriptsize}
\setlength{\tabcolsep}{4pt}

\begin{tabular}{llllll}
\hline
\\[-3pt]
\multicolumn{1}{c}{\textbf{Prior}} & \multicolumn{1}{c}{\textbf{Constraint}} & \multicolumn{1}{c}{\textbf{LOOIC}} & \multicolumn{1}{c}{\textbf{Standard Error}} & \multicolumn{1}{c}{\textbf{WAIC}} & \multicolumn{1}{c}{\textbf{Standard Error}}\\
 \\[2pt]
\hline
\\[-3pt]
SB2(0.5, 0.5, 100) & Original    & 41,900.4   & 5,572.9  & 62,587.9 & 10,326.5 \\[2pt]
SB2(0.5, 0.5, 100) & Sum-to-zero & 41,926.1   & 5,581.7  & 62,766.3 & 10,332.3\\[2pt]
G(0.001, 0.001)    & Original    & 41,919.8   & 5,591.9  & 62,837.0 & 10,438.8\\[2pt]
G(0.001, 0.001)    & Sum-to-zero & 41,986.0   & 5,587.4  & 62,543.6 & 10,349.0\\[2pt]
 \\[3pt]
	\hline
\end{tabular}
\end{scriptsize}
\end{table}

Table \ref{tab:tfr} shows the fitted TFR median values in the 2018-2022 period for different prior selections. When compared to the Gamma prior, all Scaled Beta2 priors estimate a TFR closer to 0.9, the actual TFR in the 2018-2022 period. The Scaled Beta2 models produce reasonable credible intervals, while the Gamma model has a very wide credible interval that does allow not a practical interpretation. 

\begin{table}[H]
  \centering
  \caption{\label{tab:tfr}Comparison of fitted TFR for the 2018-2022 period through precision prior types and effect constraints.}
  \begin{scriptsize}
  \begin{tabular}{llll}
\hline
\\[-3pt]
     \textbf{Model prior} & \textbf{Fitted TFR Median, 2018-2022} & \textbf{95\% lower limit} & \textbf{95\% upper limit} \\
\\[2pt]
\hline
\\[-3pt]
     SB2(0.5, 0.5, 100) & 1.07 & 0.87 & 1.30 \\
     SB2(1, 5, 5) & 1.07 & 0.77 & 1.58 \\
     SB2(2, 5, 6.25) & 1.06 & 0.75 & 1.54 \\
     SB2(1, 25, 25) & 1.06 & 0.56 & 2.06 \\
     SB2(2, 25, 156.25) & 1.07 & 0.77 & 1.46 \\
     Gamma(0.001, 0.001), original constraint & 1.14 & 0.15 & 9.07 \\
     Gamma(0.001, 0.001, sum-to-zero constraint & 1.18 & 0.15 & 9.27
 \\[3pt]
	\hline
   \end{tabular}
   
  \end{scriptsize}
\end{table}

\section*{Discussion}

All criteria used for model comparison lead us to conclude that cohort effects have greater weight over period effects when analyzing fertility decline in Puerto Rico. As observed in both the explanatory and statistical analyses, birth rates are highest among women aged 20-24, (see Figure \ref{eff-plt}). A postponement of births is thus not observed in Puerto Rico, making the process of fertility recovery unlikely. This lack of postponement is consistent with the results in \cite{murray2018population}, where Puerto Rico had the lowest Total Fertility Rate for women over 30 years of age (0.3, UI 0.29-0.41) in the world, in a comparative study of 195 countries. The TFO30 is defined in the article as the number of livebirths expected for a hypothetical woman ageing from 30 to 54 years. When observing the period effects in Figure \ref{eff-plt}, the decrease seems to coincide with the decline in the number of women in reproductive age, that started in the 1998-2002 period. Puerto Rican fertility shows a different pattern to countries such as South Korea, United States, Italy and Singapore, by depending more on cohort effects rather than period effects. 
While lowest-low fertility is mostly attributed to fertility postponement
\citep{goldstein2009}, Puerto Rico does not experience postponement of births, reinforcing the need to consider alternate factors that contribute to low fertility levels in the island. 

According to Bhrolchain \cite{bhrolchain1992}, considering previous history could imply a cohort approach provided that each cohort’s reproductive behavior is persistently different across its reproductive history. This notion is supported by the results shown in Figure 3, the model effects and the lack of postponement, showing different reproductive patterns across cohorts. We emphasize that our study is consistent with the lack of postponement of births in Puerto Rico previously observed in \citep{murray2018population}. Puerto Rico’s strong cohort predominance is rare. Other countries show either period dominance or a combination of period and cohort patterns. For Nordic countries, cohorts born in the late 1980s show a notable quantum decline (ie., younger mothers have smaller families), \citep{hellstrand2021not}, while earlier cohorts have an evident postponement of births with substantial recuperation \citep{andersson2009cohort}. In Spain, differences in age specific fertility among cohorts are correlated with women's education and labor-market conditions, \citep{ahn2020analysis} which is a classic example of period-related factors. In East Asia, studies on China and Taiwan \citep{lan2021, tzeng2019trends} show that cohort related dynamics are linked to public policy, such as China’s one-child policy. As in Taiwanese fertility, recent studies on South Korean \citep{hwang2023} and Japanese \citep{okui2020marriage} fertility emphasize marriage timing as well, which is associated with the cohort effect, even when period effects have greater weight overall. 

The differences among period effects in the APC and AP models, also give importance to cohort effects. From Figure \ref{apc-applot}, there is a greater gap in more recent periods, suggesting that cohort changes have influenced birth rates in recent years. Both the residual sums of squares and the area criterion suggest that cohort effects are dominant when explaining changes in fertility.

The LOOIC favors a Scaled Beta2 prior, and the use of our original constraints. Using a Gamma prior with a sum-to-zero constraint is preferred by the WAIC, however, regardless of the type of constraint used, the Gamma models do not generate appropriate TFR estimates. As described in \cite{vehtari2017} the LOOIC is preferable over the WAIC for its robustness. The use of the LOOIC over the WAIC is supported by our model results, since the LOOIC coincides with the results found in the residual sum of squares criterion and the fitted TFR criterion. This stresses the need to use multiple model comparison criteria for model selection and interpretation. We conclude that overall, the Scaled Beta2 is more suitable, since it fits the data better. The different model comparison criteria support the use of our original constraints.

\section*{Conclusion}

Contrary to the rigid specification of classical APC implementations in R, such as the \textit{apc} package, our Bayesian model permits a novel way to better understand, specify, and address the model restrictions for the identification problem.  The model implementation also permits the test of different prior distributions for the scale parameters, showing that the Scaled Beta2 distribution represents a good alternative to the criticized \citep{gelman2006prior, perez2017scaled} Inverse Gamma/Gamma distribution. While the identification problem cannot be fully solved, our alternative of imposing constraints on completed cohorts is an improvement on the traditional equality constraint, as well as the sum-to-zero constraint, as was shown in Tables \ref{tab:const} and \ref{tab:tfr}. Our alternative methodology, which includes applying the Scaled Beta2 distribution, could potentially be very useful when studying other jurisdictions and territories that present low fertility levels. These methodological contributions could also be adapted to the development of Bayesian panel data models. Moreover, our framework facilitates the interpretation of the APC effects through the proposed alternative model criteria. Regarding demographic methodologies, we stress the importance of considering each territory’s unique demographic context when studying fertility in lowest-low situations, instead of solely relying on existing period approaches. Considering the cohort effects alongside the period framework will lead to a comprehensive analysis. 

As future work, Bayes factors will be developed as additional model criteria for APC models. Bayes factors are a better alternative for hypothesis testing, which is necessary to evaluate the weights of period and cohort effects in APC models in a robust matter. 

\section*{Supplementary Information}

Coding and datasets for the replication with R of all paper analyses are availaInformation on model implementation and datasets can be available upon request (jomarie.jimenez@upr.edu).  A GitHub repository with the model code will be available at the moment of publication.

\section*{Acknowledgments}
This research is supported by Institutional Funds of the University of Puerto Rico for the Demography Study of Puerto Rico (2024 and 2026).

\newpage

\addcontentsline{toc}{section}{\numberline{}References}
\bibliography{refs.bib} 
\nocite{*}

\end{document}